\documentstyle[aps,epsfig,twocolumn]{revtex}

\begin{document}

\title{Vortex dynamics, pinning, and critical currents 
in a Ginzburg-Landau type-II superconductor}

\author{T. Winiecki and C. S. Adams}

\address{Department of Physics, University of Durham , Durham DH1 3LE.}

\date{\today}
\draft
\maketitle
\begin{abstract}

The dynamics of vortices in a type-II superconductor with defects are studied 
by solving the time-dependent Ginzburg-Landau equations in two and three
dimensions. We show that vortex flux tubes are trapped by volume defects 
up to a critical current density where they begin to jump between pinning sites along
static flow channels. We study the dependence of the critical current on the
pinning distribution and find for random distributions a maximum
critical current equal to a few percent of the depairing current
at a pinning density three times larger than the vortex 
line density. Whereas for a regular triangular pinning array, the critical
current is significantly larger when the pinning density matches 
the vortex line density.

\end{abstract}


In a type-II superconductor, dissipation is associated with the motion of the 
vortex lattice \cite{tink96,kim65}. This dissipation is reduced by the presence 
of defects, which pin the vortex lattice up to a critical current density
where depinning occurs. In many applications such as 
superconducting magnets, one is interested 
in optimizing the vortex pinning to achieve the maximum critical current. However,
the details of the depinning transition are complex involving the non-equilibrium 
dynamics of an elastic lattice through a disordered medium. Theoretical
studies based on molecular dynamics simulations suggest the existence of 
various dynamical phases of vortex motion including plastic flow, 
uncoupled static channels and coupled channels \cite{moon96,olso98}. 
It is also possible to simulate vortex dynamics by solving the time-dependent 
Ginzburg-Landau equations \cite{kato93,mach93,mach94}, where the 
vortex-vortex interaction is completely characterised by the Ginzburg-Landau 
parameter, $\kappa$. However, three dimensional Ginzburg-Landau
vortex dynamics simulations are computationally
intensive, in part because the standard explicit integration
methods require very small time-steps. In contrast, semi-implicit
methods are second order accurate in time allowing 
large time-steps. Although semi-implicit methods
are widely used to simulate three dimensional vortex dynamics in dilute
Bose-Einstein condensates \cite{wini00}, 
the Ginzburg-Landau equations, involving coupled time-dependent
vector fields, are more complex.

In this paper we develop a semi-implicit method to solve the time-dependent 
Ginzburg-Landau equations in three dimensions \cite{wini01}. 
For intermediate values of $\kappa$, the semi-implicit 
method is two orders of magnitude faster than explicit methods, 
making it feasible to study dynamical vortex phases, 
depinning, and the dependence of the critical current on the density and 
distribution of pinning sites. Although pinning may arise due
to magnetic defects, dislocations, grain boundaries, and correlated
disorder such as twin planes in high-$T_c$ superconductors,
we restrict the current study to volume defects which exclude the supercurrent.

The time-dependent Ginzburg-Landau equations can be written as
\begin{eqnarray}
\partial_t\psi & = & (\nabla-{\rm i}A)^2\psi-(\vert\psi\vert^2-1)\psi\\
\partial_t A & = & (\nabla S -A)\vert\psi\vert^2 -\kappa^2\nabla\times\nabla\times A
\end{eqnarray}
where $\psi$ is the order parameter, $A$ is the vector potential, 
$S$ is the phase of $\psi$, and $\kappa=\lambda/\xi$, where 
$\lambda$ and $\xi$ are the penetration depth and
coherence length, respectively. In equations (1) and 
(2), distance is measured in terms of 
$\xi$, time in terms of relaxation time, $\tau=\xi^2/D$, where 
$D$ is the diffusion constant, and the magnetic field in terms 
of the upper critical field, $H_{\rm c2}$. In addition, the Meissner state critical
field is given by $H_{\rm c}=1/\sqrt{2}\kappa$, 
and the depairing current density by $j_D = 2/3\sqrt{3}=0.385$ 
\cite{tink96}. The equations are discretized using a grid of 
$51\times51\times 51$ points 
with a grid spacing $h=0.4$. The gauge invariance
of the discretized equations is preserved by introducing link variables 
of the form $U^x_{ijk}=\exp(-{\rm i}A^x_{ijk}h)$ \cite{grop96}.  
The discretized equations are solved using a semi-implicit Crank-Nicholson
method with a time step, $\delta t=0.5$ \cite{wini01}.
A current flow along $x$ is 
induced by imposing a magnetic field difference, $\Delta B_z$, between
the upper ($y=10$) and lower ($y=-10$) 
boundaries. The supercurrent across the boundary is set to zero. 
We impose periodic boundary conditions at $x=\pm10$ and $z=\pm10$.
The average current density is given by, 
$j=\kappa^2 \Delta B_z/d$, where $d$ is the width of 
the superconductor. 
A pinning array is produced by adding a potential term to equation 
(1) consisting of a random distribution of cubic potential
steps with side length $a=1.2$ and height $V_0=5.0$. 
In agreement with other studies \cite{stej92}, 
we find that the pinning strength increases with $a$ for $a<\xi$, and 
saturates for $a>\xi$. A more
sophisticated pinning model would be needed to account for the 
large pinning forces observed for small defects \cite{thun82}.

Fig.~1 shows a sequence of images illustrating 
the motion of the vortex lattice through the pinning array. 
 In frame 1, six flux tubes are visible.
By comparing frames 1, 2 and 3, one sees that the central flux tubes are moving
whereas the two pairs on either side are pinned. However,
between frames 4 and 5 the flux tubes on the left and
right jump to the next pinning site. This differential motion between
neighbouring planes in the vortex lattice plays an important
role in the voltage-current
characteristic (see below). After frame 6, a similar
but not identical sequence recurs. For the simulations presented
in Fig.~1, the bending of the vortex lines
is increased by the choice of a larger value of $\kappa$ and strong pinning. 
However, no entangling of vortex lines is observed. For smaller $\kappa$,
the vortex lines become more rigid, and the behaviour of the three
dimensional system and a two dimensional cross section 
are qualitatively very similar. For high-$T_c$ superconductors,
a comparison between the two and three dimensional dynamics should consider
possible effects of the layered structure \cite{blat94}. 

\begin{figure}[hbt]
\centering
\epsfig{file=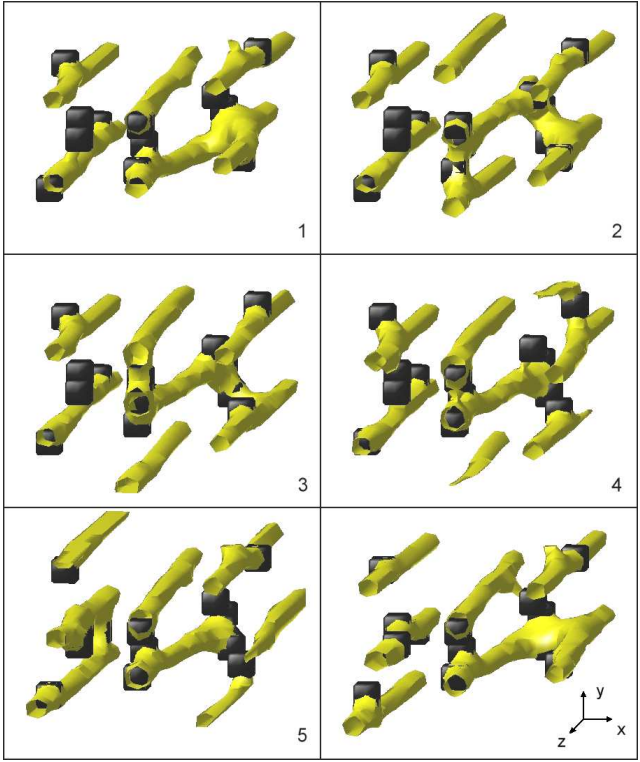,width=8.5cm,clip=,bbllx=40,bblly=30,bburx=570,bbury=674}
\caption{
A sequence of three dimensional images showing the motion of a 
$\kappa=5$ vortex lattice through a random pinning array. 
The axes are shown inset in frame 6. The current flows along $x$, 
the external magnetic field is 
along $z$, and the vortices move in the $-y$ direction. Each frame shows 
a region with dimensions $9\times7\times20$ coherence lengths containing 12 
pinning sites (shown in black, not to scale). The external magnetic field
and current are $B_{\rm ext}=0.4$ and $j=5\times10^{-3}$, respectively. The grey 
flux tubes corresponds to surfaces of constant supercurrent density,
$\vert\psi\vert^2=0.05$. The time interval between successive frames is 100.
}
\label{fig:1}
\end{figure}

We use two dimensional simulations to study the effect of pinning on the 
voltage-current characteristic or $V-I$ curve of a superconductor with
$\kappa=3$, where three dimensional effects are suppressed. In addition, we reduce the 
size and strength of the pinning sites to $a=0.8$ and $V_0=2.0$, respectively.
In Fig. 2 we present contour plots illustrating the vortex lattice in two 
dimensions. Fig.~2(a) shows the instantaneous vortex distribution for a 
perfect superconductor (no pinning). The vortex density is 
proportional to the magnetic field which decreases linearly from the bottom to
the top. The vortices move upwards with a speed 
$v=E/B$, where $B$ is the local magnetic field and the electric field, $E$, 
is constant throughout the sample. Consequently, the vortex flow obeys a 
Bernoulli-like equation where the flow is faster in regions of lower density 
(lower magnetic field) and the dissipation can be thought of as a relaxation 
of the magnetic flux lattice.    
 
\begin{figure}[hbt]
\centering
\epsfig{file=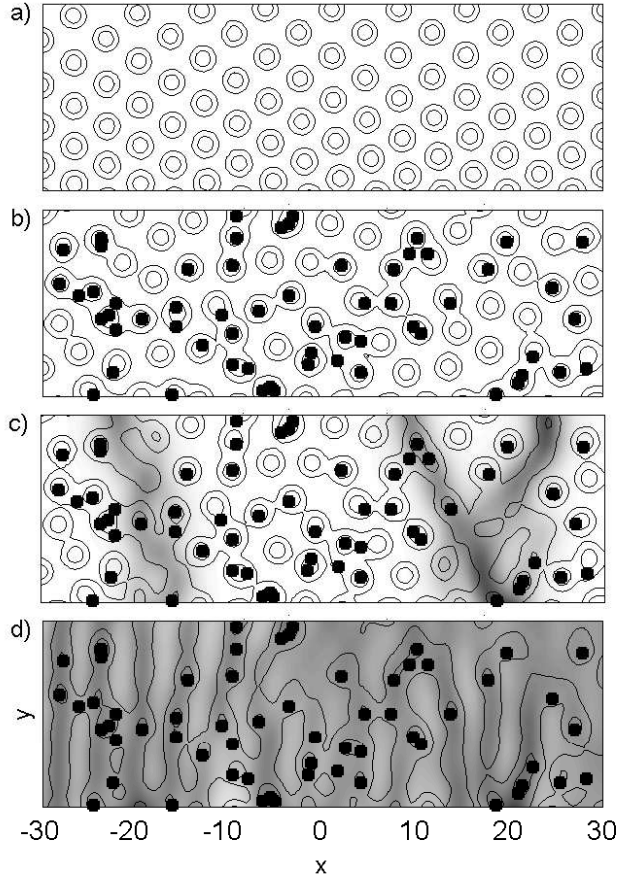,width=8.5cm,clip=,bbllx=36,bblly=180,bburx=273,bbury=510}
\caption{Contour plots of the supercurrent density for a section
of a superconductor with dimensions $60\times20$ coherence lengths,
and $\kappa=3$, subject to an external magnetic field in the $z$ direction, 
$B_{\rm ext}=0.4$. The current flows in the $x$ direction and the 
vortices move in the $y$ direction. (a) For a current $j=0.06$ and no pinning, 
the vortices form a triangular lattice with lattice spacing proportional
to the local magnetic field. (b) The addition of pinning 
(density 0.056~$\xi^{-2}$) creates a vortex glass, which at low currents, 
$j=0.02$, is pinned. (c) At intermediate currents, 
$j=0.025$, vortex motion begins along channels, indicated by the grey 
scale image of the local electric field, superimposed on a 
time-averaged contour plot of the supercurrent density. (d) At larger 
currents, $j=0.054$, all the vortices are moving  
and the electric field is non-zero everywhere, however, 
the channels, where vortex motion mainly occurs, 
are still visible.}
\label{fig:2}
\end{figure}

Adding defects transforms the triangular lattice into an irregular
vortex glass, Fig.~2(b). For low driving fields, the vortex
glass is frozen. As the current is increased, individual
vortices begin to jump between pinning sites. As in the
three dimensional simulations, Fig.~1, this motion begins along
channels. The existence of static channels confirms the results of 
molecular dynamics simulations \cite{moon96,olso98}. In the 
Ginzburg-Landau model channels can merge or divide at
intermediate drive currents, as shown in Fig. 2(c). At larger currents,
all the vortices are moving but the channels are still evident,
Fig.~2(d). 

The on-set of vortex motion coincides with the on-set of dissipation or 
breakdown of superconductivity. In Fig.~3 we plot the $V-I$ curve for a 
two-dimensional thin film for different defect 
densities. The voltage is measured by decreasing the current at a very 
slow rate of $-1.2\times10^{-7}$ in $2.5\times10^{5}$ steps,
and the $V-I$ curves is obtained from a 200 point moving average. As our sample
size is relatively small, surface effects tends to dominate. The 
critical current due to the Bean-Livingston barrier for vortices entering 
and leaving the calculation region \cite{kato93} is the same order 
of magnitude as the pinning effect. 
In order to study pinning only we remove the surface effects by 
adding a boundary layer 
of width 9 $\xi$ on either side of a calculation region with width 30 $\xi$.
Within the boundary layer, a linear ramp potential 
reduces the supercurrent density gradually to zero. The current density 
and the voltage are measured 
within the calculation region ($\vert y\vert\leq15$) only. 

\begin{figure}[hbt]
\centering
\epsfig{file=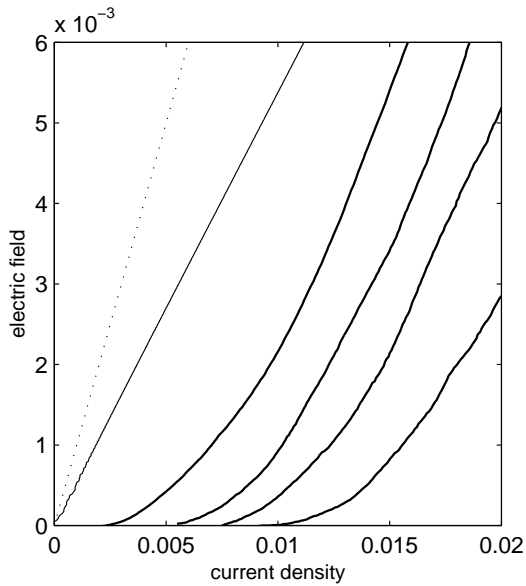,width=7cm,clip=,bbllx=45,bblly=175,bburx=275,bbury=435}
\caption{The $V-I$ curves for a two dimensional section of superconductor with
pinning densities (from the right) 0.14, 0.28, 0.39, and 0.56~$\xi^{-2}$
(at these relatively high densities, the critical current decreases with 
increasing pinning density). The thin black line corresponds to the $V-I$ curve 
without pinning, and the dotted line shows the normal resistance, $E=j$. 
Note that at large currents the slope of the $V-I$ curves 
is similar to the normal resistance curve.}
\label{fig:3}
\end{figure}

The shape of the $V-I$ 
curve is dependent on the details of the vortex dynamics. The characteristic 
`curved foot' can be explained by the combination of an increase in the 
number of vortex flow channels and increased flow along each active channel, 
as illustrated in Fig. 2(c) and (d). The $V-I$ curve becomes linear when all 
the vortices start to move. 
The ratio between the $V-I$ curves and the normal resistance (the dotted 
line in Fig.~3) gives the dimensionless 
resistivity, which measures the fraction of current carried by normal electrons.

As the current is decreased the voltage
becomes zero, i.e., all the vortices become pinned, at some finite current
which we define as the critical current density, $j_c$. 
In the absence of finite temperature induced fluctuations or
vortex creep, the value of $j_c$ is well defined. 
However the critical current is sensitive to the exact distribution of pinning sites, 
therefore we average 
over six random distributions with the same density.
Fig. 4 shows a plot of the average value of $j_c$ against pinning density. 
The maximum critical current density is about 2~$\%$ of the depairing
current, $j_D$. For comparison, the optimum critical current
density of Nb-Ti alloy is $\sim3$~$\%$ of $j_D$. 
The maximum value of $j_c$ occurs at pinning density about three times 
larger than the vortex line density (indicated by the dotted line in Fig. 4).
  
\begin{figure}[hbt]
\centering
\epsfig{file=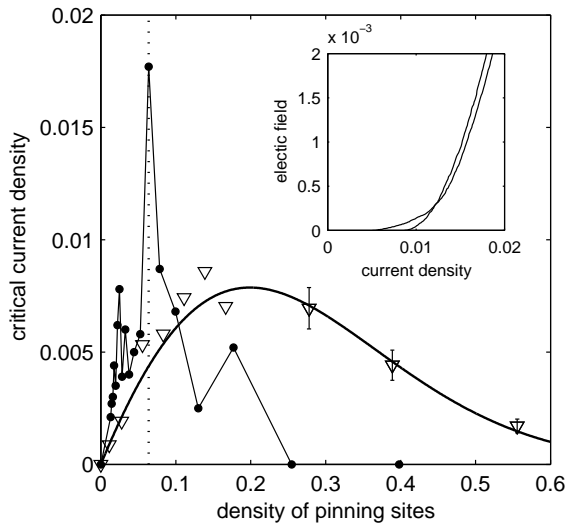,width=7.5cm,clip=,bbllx=18,bblly=214,bburx=253,bbury=434}
\caption{The critical current density as a function of the defect density
(in units of $\xi^{-2}$) for both random distributions ($\bigtriangledown$) and 
regular triangular arrays ($\bullet$). The data points are 
determined from an average of six random distributions. The error
bars (shown for the high density distribution only) indicate the standard 
deviation. An example illustrating the effect of the distribution
on the $V-I$ curves is shown inset. The bold curve is a fit 
using the function $Ax\exp(-Bx)$, where $A$ and $B$
are fit parameters. The critical current density for a regular
triangular array is a maximum when the pinning density is equal to
vortex line density (indicated by the dotted line). }
\label{fig:4}
\end{figure}

The dependence of the critical current on the defect density
fits reasonably well to a function of the form $Ax\exp(-Bx)$.
The linear increase at low pinning density follows from the
linear dependence of the critical current on the pinning
force. The exponential decrease at large pinning densities is due to
the competing effect of supercurrent depletion by defects.
The shape of the curve and the relatively high optimum pinning
density also agree qualitatively with experimental results
on silver doped high-$T_c$ superconductors \cite{savv91}.
 
For certain random distributions one finds persistence static channels which can
dramatically reduce the critical current. This is
illustrated in Fig.~4(inset), where the curve with lower
dissipation at large currents has a significantly lower critical current.
 
One approach to increase the critical current is to introduce
a regular pinning array by nanostructuring \cite{mosh98,vanb00,morg01}.
In Fig.~4 we show that a regular triangular array increases
the critical currents by more than a factor of two, however, the
optimum pinning density is sharply peaked around the vortex line
density. Consequently, the enhancement is only obtained
within a narrow range of the external magnetic field. This agrees with 
experimental studies where a sharp enhancement peak is obtained
at matching magnetic field values \cite{morg01}.
There are two additional critical current peaks, one at one third the 
vortex line density where every third vortex is 
trapped, and one at half the vortex line density, which is
weaker because the matching only occurs on alternate planes.
For small pinning sites ($a=0.8$ compared to the vortex cores size 
of 2) the maximum critical current is about 5~$\%$ of the depairing current, $j_D$.
For $a=2$ we obtain $j_c=0.074j_D$, which
suggests that other pinning mechanisms may be needed to
obtain $j_c \sim j_D$. 

In summary, we have studied vortex dynamics and pinning in a three dimensional 
superconductor by solving the time-dependent Ginzburg-Landau
equations. We find that
above a critical current density vortex flux tubes jump
between pinning sites following specific channels.
The main features of the dynamics are reproduced 
by two dimensional simulations. We study the effect of pinning 
on the voltage-current characteristic of the superconductor,
and show that the breakdown of superconductivity is associated
with the appearance of channelled vortex flow. The 
characteristic curved foot in the $V-I$ curve arises due to
the combination of the formation of more channels and
faster vortex flow along each channel. For a random pinning array
we find a maximum critical current equal to 2~$\%$ of the depairing
current occurring at a pinning density of about three times the vortex 
line density. Finally, we study the critical currents produced by vortex 
matching pinning arrays. The results suggest that time-dependent 
Ginzburg-Landau simulations are ideally
suited to provide quantitative predictions of critical currents in 
type-II superconductors.

\acknowledgements
We thank S. J. Bending for comments. Financial support was provided by the EPSRC.

\end{document}